\documentclass[12pt]{article}

\usepackage{amssymb}
\let\a=\alpha

      \let\x=\xi           
        \let\f=\phi
         \let\w=\omega
\newcommand{\beq}{\begin{equation}}
\newcommand{\eeq}{\end{equation}}
\newcommand{\beqn}{\begin{eqnarray}}
\newcommand{\eeqn}{\end{eqnarray}}

\newcommand{\nn}{\nonumber}

\newcommand{\be}{\begin{equation}}
\newcommand{\ee}{\end{equation}}
\newcommand{\ba}{\begin{eqnarray}}
\newcommand{\ea}{\end{eqnarray}}
\newcommand{\bdm}{\begin{displaymath}}
\newcommand{\edm}{\end{displaymath}}

\def\a{\alpha}

\newcommand{\ie}{{\it i.e.\ }}
\newcommand{\eg}{{\it e.g.\ }}

\DeclareMathAlphabet{\mathpzc}{OT1}{pzc}{m}{it}
\def\bea{\begin{eqnarray}}
\def\eea{\end{eqnarray}}
\def\beas{\begin{eqnarray*}}
\def\eeas{\end{eqnarray*}}
\def\sla{\raise.15ex\hbox{$/$}\kern-.57em}

\def\bea{\begin{eqnarray}}
\def\eea{\end{eqnarray}}
\def\de{\partial}

\def\sla{\raise.15ex\hbox{$/$}\kern-.57em}
\def\ie{{\it i.e.}~}
\def\eg{{\it e.g.}~}

\def\a{\alpha}

\def\th{\theta}

\def\f{\phi}

\def\w{\omega}
\def\vf{\varphi}

\def\cA{{\cal A}}
\def\cB{{\cal B}}

\def\cF{{\cal F}}
\def\cG{{\cal G}}

\def\cK{{\cal K}}
\def\cL{{\cal L}}
\def\cM{{\cal M}}
\def\cN{{\cal N}}

\def\cV{{\cal V}}

\def\cY{{\cal Y}}

\begin{document}
\begin{titlepage}
\begin{flushright}
{ROM2F/2009/33}\\
\end{flushright}
%
%
\begin{center}
{\Large\bf On Exact Symmetries and Massless Vectors  \vskip 0.1cm
in Holographic Flows and other Flux Vacua} \\
\end{center}
\begin{center}
{\bf Massimo Bianchi$^{1,2}$, Wayne de Paula$^{1,3}$} \vskip 0.3cm
{\sl $^{1}$ Dipartimento di Fisica, Universit\`a di Roma ``Tor Vergata''\\
 I.N.F.N. Sezione di Roma ``Tor Vergata''\\
Via della Ricerca Scientifica, 00133 Roma, Italy}\\
 {\sl $^{2}$ Physics Department, Theory Unit, CERN \\ CH 1211,
Geneva 23, Switzerland}\\
{\sl $^{3}$ Departamento de Fisica, Instituto Tecnologico de Aeronautica, \\
12228-900 Sao Jose dos Campos, SP, Brazil}
\end{center}
\vskip 1.0cm
\begin{center}
{\large \bf Abstract}
\end{center}
We analyze the isometries of Type IIB flux vacua based on the
Papadopolous-Tseytlin ansatz and identify the related massless
bulk vector fields. To this end we devise a general ansatz, valid
in any flux compactification, for the fluctuations of the metric
and $p$-forms that diagonalizes the coupled equations. We then
illustrate the procedure in the simple case of holographic flows
driven by the RR 3-form flux only. Specifically we study the fate
of the isometries of the Maldacena-Nu${\rm\tilde{n}}$ez solution
associated to wrapped D5-branes.



\vfill

\end{titlepage}


\section{Introduction}

Flux vacua \cite{Giddings:2001yu} find many interesting
applications in String Theory, ranging from holographic flows dual
to non (super)conformal boundary theories \cite{Klebanov:2000nc,
Klebanov:2000hb, Maldacena:2000yy, Papadopoulos:2000gj,
Apreda:2001qb, Herzog:2002ih, Apreda:2003sy} to moduli
stabilization in phenomenologically viable models with open and
unoriented strings and otherwise \cite{Bianchi:2009va}. More
recently unexpected implications of the AdS/CFT correspondence
\cite{Maldacena:1997re, Gubser:1998bc, Witten:1998qj,
Gubser:2009md}  in condensed matter physics \cite{Hartnoll:2009sz,
Bhattacharyya:2008kq}, astrophysics \cite{Hartman:2009nz,
deBoer:2009wk} and gravity at a Lifschitz point
\cite{Horava:2009uw} have attracted a lot of attention.

Due to the presence of fluxes, fields belonging to different
sectors tend to mix with one another, compatibly with the residual
(super)symmetry. Resolving the mixing and finding the spectrum of
excitations is extremely laborious \cite{Bianchi:2000sm,
Bianchi:2001de, Bianchi:2001kw} as witnessed by the enormous
effort needed to accomplish the task for the metric and
active\footnote{By {\it active} we indicate those scalars with a
non trivial profile in the background. All the other
(pseudo)scalars are said to be {\it inert}.} scalar modes in
holographic flows described by the Papadopoulos-Tseytlin (PT)
ansatz \cite{Krasnitz:2000ir, Krasnitz:2002ct, Berg:2005pd,
Berg:2006xy}. Our aim is to extend this kind of analysis to the
vector sector \cite{Bianchi:2000sm, Bianchi:2001de,
Bianchi:2001kw, Brandhuber:2000fr, Brandhuber:2002rx}. To this
end, we will start by studying the fate of bulk symmetries of the
Type IIB supergravity solutions. Although we will mostly adopt a
10-d perspective, we will also present the 5-d viewpoint, that has
a more direct applicability in Holographic Renormalization
\cite{de Haro:2000xn, Bianchi:2001de, Bianchi:2001kw,
Kalkkinen:2001vg, Martelli:2002sp, Bianchi:2003bd, Bianchi:2003ug,
Mueck:2004qg, Papadimitriou:2004rz} and Holographic QCD
\cite{Erdmenger:2007cm, Kiritsis:2009hu, Gursoy:2009jd,
dePaula:2008fp, dePaula:2009za}.

Symmetries can be divided into three classes \begin{itemize}
\item{Exact Symmetries: not only the metric admits Killing vectors
but also fluxes are invariant} \cite{Bianchi:2000sm,
Bianchi:2001de, Bianchi:2001kw, Brandhuber:2000fr,
Brandhuber:2002rx}. \item{Partially Broken Symmetries: Metric
invariant, some fluxes are not} \item{Broken Symmetries: Metric
and fluxes only asymptotically invariant} \cite{Krasnitz:2000ir,
Krasnitz:2002ct, Klebanov:2002gr, Gubser:2004qj}
\end{itemize}

The PT ansatz \cite{Papadopoulos:2000gj} enjoys $SU(2)\times
\widetilde{SU(2)}$ isometry for arbitrary choices of the `radial'
functions. On the contrary, the $U(1)_R$, associated to shifts of
the coordinate $\psi$, is broken except for very special cases.
The breaking is spontaneous from the bulk viewpoint, \ie the
would-be massless vector field becomes massive after `eating' a
Goldstone boson. The St\"uckelberg formalism for the gauging of
axionic shift symmetries is particularly convenient in this
respect  \cite{Bianchi:2000sm, Bianchi:2001de, Bianchi:2001kw,
Brandhuber:2000fr, Brandhuber:2002rx, Krasnitz:2000ir,
Krasnitz:2002ct, Klebanov:2002gr, Gubser:2004qj}. Except for some
very general remarks, we will neither have much further to say
about broken symmetries and massive vectors nor discuss at all
massless vectors related to harmonic forms \cite{Herzog:2002ih}
and to probe branes \cite{Erdmenger:2007cm}. The latter give rise
to chiral `flavor' symmetries (breaking) and mesons. The former to
baryonic symmetries.

The plan of the paper is as follows. In section 2 we describe the
5-d Lagrangian governing the dynamics of gauge fields and their
mixing with would-be axions. Next, in Section 3, we check that the
PT ansatz indeed admits full $SU(2)\times \widetilde{SU(2)}$
symmetry, in that not only the metric but also the other
background field (strengths) are invariant. In Section 4, we
identify the bulk vector fields that remain massless by means of
an ansatz for the fluctuations of the metric and $p$-forms, that
diagonalizes the coupled equations. Finally, in Section 5 we
illustrate the procedure in the simple case of holographic flows
and other flux vacua with $F_3$ only, which are invariant under
$\Omega$. Specifically we study the MN solution
\cite{Maldacena:2000yy} associated to wrapped D5-branes, \ie
`fractional' D3-branes. Our conclusions and summary are contained
in Section 6.

\section{Vector fields in Holographic Renormalization}

The 5-d Lagrangian describing vector fields and their possible
mixing with inert (pseudo)scalars \eg axions reads
\cite{Bianchi:2000sm, Bianchi:2001de, Bianchi:2001kw,
Brandhuber:2000fr, Brandhuber:2002rx} \be \cL_{v-a} = -{1\over 4}
\cK_{ij} F_{\mu\nu}^i F^{\mu\nu j} + {1\over 2} h_{AB}
(\de_\mu\beta^A - M^A_i A_\mu^i)(\de^\mu\beta^B - M^B_j A^{\mu j})
\ee The gauge kinetic function $\cK_{ij}$ and the axion metric $
h_{AB}$ may depend on the (active) scalars, while, in the present
parametrization $M^A_i$ are constant mass parameters\footnote{In
general, after gauging some isometry of the scalar metric
$\cG_{ab}(\phi)$, covariant derivatives are given by $D\phi^a =
\de \phi^a + \cK^a_i(\phi) A^i$, where $\cK^a_i$ are Killing
vectors \ie $\nabla^a_{\cG} \cK^b_i(\phi) + \nabla^b_{\cG}
\cK^a_i(\phi) = 0, \forall{} i$. Here we focus on the gauging of
axionic shifts $\delta\beta^A = M^A_i \alpha^i_0$.}. The
Lagrangian is invariant under gauge transformations of the form
\be \delta A^i = \de\alpha^i \quad , \quad \delta\beta^A = M^A_i
\alpha^i. \ee The square mass matrix \be \cM^2_{ij}(\phi) =
h_{AB}M^A_iM^B_j \ee is semi-positive definite. Zero eigenvalues
correspond to exactly massless vectors, associated to isometries
or to harmonic forms present in the background solution. Non-zero
eigenvalues correspond to massive vectors and broken symmetries.
Introducing gauge invariant combinations for the latter \be
\cB^{\hat{i}}_\mu = A^{\hat{i}}_\mu - (\cM^{-2}_{\perp
0})^{{\hat{i}}{\hat{j}}}M_{\hat{j}}^A h_{AB}\de_\mu\beta^B \ee and
denoting by $A_\mu^{i_o}$ the former yields \be \cL_{v} = -{1\over
4} \cK_{i_oj_o}(\phi) F_{\mu\nu}^{i_o} F^{\mu\nu j_o} -{1\over 4}
\cK_{\hat{i}\hat{j}}(\phi) \cF_{\mu\nu}^{\hat{i}} \cF^{\mu\nu
\hat{j}} + {1\over 2} \cM^2_{\hat{i}\hat{j}}(\phi)
\cB_\mu^{\hat{i}}\cB^{\mu \hat{j}} \ee After diagonalization, one
finds a collection of decoupled vector bosons\footnote{At the
quadratic level this is true for non-abelian symmetries, too.}
each described by \be \cL_{v} = -{1\over 4} \cK(\phi) \cF_{\mu\nu}
\cF^{\mu\nu } + {1\over 2} \cM^2(\phi)\cA_\mu\cA^\mu \ee where
$\cK(\phi)$ is the resulting gauge kinetic function and
$\cM(\phi)$ is the possibly vanishing mass.

Putting the kinetic term in canonical form one has
\cite{Bianchi:2000sm} \be \cM^2_{eff} = {1\over 2} {\cK''\over
\cK} +  {A'\cK'\over \cK} - {1\over 4} \left({\cK'\over
\cK}\right)^2 + {\cM^2\over \cK}  \ee where primes denote
derivatives wrt the holographic radial variable. Clearly $\cM^2
=0$ for bulk vector fields associated to unbroken boundary
currents (exact global symmeries), while $\cM^2 \neq 0$ for broken
symmetries. In most if not all known cases \cite{Bianchi:2000sm}
\be \cM^2_{eff} = - 2 A'' \ee Quite remarkably but without a clear
explanation, the above relation has been verified for (transverse)
vectors fields in all known solutions: Coulomb branch flow with
$SO(6)\rightarrow SO(4)\times SO(2)$ \cite{Bianchi:2000sm,
Bianchi:2001de, Bianchi:2001kw, Brandhuber:2000fr,
Brandhuber:2002rx}, GPPZ flow $SO(6)\rightarrow SO(3)\times
U(1)_R$ \cite{Bianchi:2000sm, Bianchi:2001de, Bianchi:2001kw}, KT
(and partially KS) solution with (broken) $U(1)_R$ R-symmetry
\cite{Krasnitz:2000ir, Krasnitz:2002ct}.

\section{Field equations and PT ansatz}

To set the stage for our analysis, let us now briefly recall Papadopoulos and Tseytlin  (PT) ansatz for flux vacua in Type IIB supergravity and
its symmetries. The main motivation behind PT ansatz is to identify a subset of fields that form a consistent truncation of Type IIB supergravity
and allow to study flux vacua with reduced or no supersymmetry at all. The reader familiar with Type IIB supergravity and the PT ansatz can skip
the following part and go directly to Section 3.2.

In the Einstein frame, Type IIB supergravity equations read \bea
 &&R_{MN} = {1\over 2} \de_M \phi \de_N \phi + {1\over 2} e^{2\phi} \de_M \chi \de_N \chi
 +{1\over 96}
 \hat{G}_{MPQKL}\hat{G}_N{}^{PQKL} + \\ &&{1\over 4} e^\phi
 \hat{F}_{MPQ}\hat{F}_N{}^{PQ} + {1\over 4} e^{-\phi}
 H_{MPQ}H_N{}^{PQ} - {1\over 48} g_{MN}
[e^\phi  \hat{F}_{LPQ}\hat{F}^{LPQ} + e^{-\phi} H_{LPQ}H^{LPQ}] \nn \\
 &&\nabla^2\phi  = e^{2\phi} \de_M \chi \de^M \chi
 + {1\over 12}
e^\phi  \hat{F}_{LMN}\hat{F}^{LMN} - {1\over 12} e^{-\phi} H_{LMN}H^{LMN} \\
&&\nabla^M(e^{2\phi}\de_M\chi)  = - {1\over 6} e^{\phi} H_{LMN}F^{LMN}  \\
&&\nabla^M(e^{\phi}\hat{F}_{MNP})  = {1\over 6} {G}_{NPQRS}H^{QRS}  \\
&&\nabla^M(e^{-\phi}H_{MNP} - e^{\phi}\chi \hat{F}_{MNP})
= - {1\over 6} {G}_{NPQRS}{F}^{QRS}   \\
&&\hat{G}_{M_1...M_5} = {1\over
120}\varepsilon_{M_1...M_5M_6...M_{10}} \hat{G}^{M_6...M_{10}}
\eea where \be \hat{F}_3 = F_3- \chi H_3 \qquad \hat{F}_5 = F_5 +
B_2 F_3\ee with  \be F_1 = d\chi \qquad F_3 = dC_2 \qquad F_5 =
dA_4 \qquad H_3 = dB_2 \ee

\subsection{PT ansatz}

The consistent truncation of 10-d Type IIB supergravity found by
Papadopoulos and Tseytlin is based on the following ansatz.

\begin{itemize}

\item Metric \bea ds_{10}^2 &=& e^{2p(u)-{x}(u)}\left(e^{2A(u)}
dx\cdot dx + N_{5} du^2\right) + N_{5}[e^{{x}(u)+g(u)}\left(e_1^2 + e_2^2\right) \nonumber \\
&&+ {1\over4}e^{{x}(u)-g(u)}\left(\tilde\omega_{1}^2 +
\tilde\omega_{2}^2\right)+
{1\over4}e^{-6p(u)-{x}(u)}\tilde\omega_{3}^2]\eea where $u$
denotes the holographic radial variable, the functions $A, p, {x},
g$ depend on $u$, and the `invariant' one-forms read \bea &&e_{1}
= d\th \quad, \quad
 e_{2} = -\sin \th d\varphi \nn \\
&&\tilde\omega_{1} = \omega_{1} - a(u) e_{1} \quad, \quad
\tilde\omega_{2} = \omega_{2} - a(u) e_{2}\quad, \quad
\tilde\omega_{3} = \omega_{3} - \cot \theta e_{2} \nn \\
&& \omega_{1} = \sin \psi \sin \tilde\th d\tilde\varphi + \cos
\psi d\tilde\th \quad , \quad \omega_{2} = - \sin \psi d\tilde\th
+ \cos \psi \sin \tilde\th d\tilde\varphi \nn \\
&&\omega_{3} = d\psi + \cos \tilde\th d\tilde\varphi
 \eea

\item NS-NS dilaton and R-R axion  \be \phi = \phi(u) \quad ,
\quad \chi=0 \ee

\item NS-NS 3-form \bea H_3 &=& h_2(u) \tilde\omega_{3}{\wedge }
(\omega_{1}\wedge e_1 + \omega_{2}\wedge e_2) +
du\wedge\left[h_1'(u)\left(\omega_{1}\wedge \omega_{2} + e_1
\wedge
e_2\right)\nonumber \right.\\
&& \left. + h_{2}'(u) \left(\omega_{1}\wedge e_2 -
\omega_{2}\wedge e_1\right) + h_3'(u) \left(-\omega_{1}\wedge
\omega_{2} + e_1\wedge e_2\right)\right] \eea where $'$ denotes
derivative wrt $u$. Since $dH_3 =0$, one has $H_{3}=dB_{2}$ with
\bea B_2 &=& h_{1}(u) (e_1\wedge e_2 + \omega_{1}\wedge \omega_{2}) +
h_{2}(u) (\omega_{1}\wedge e_2 - \omega_{2}\wedge e_1) \nn \\
&&+ h_{3}(u) (-\omega_{1}\wedge \omega_{2} + e_1\wedge e_2) \eea

\item R-R 3-form \bea F_3 &=& {N_5\over 4} \left\{\tilde\omega_{3}\wedge
[(\omega_{1}\wedge \omega_{2} + e_1\wedge e_2) - b(u)
(\omega_{1}\wedge e_2 - \omega_{2}\wedge e_1)] \nonumber \right.\\
&&\left.+ b'(u) du \wedge\left(\omega_{1}\wedge e_1 +
\omega_{2}\wedge e_2\right)\right\} \eea since $dF_3 =0$, one has
$F_{3}=dC_{2}$ with \bea &&C_2 = {N_5\over 4} [\psi (e_1\wedge e_2
+ \omega_{1}\wedge \omega_{2}) + b(u) (\omega_{1}\wedge e_1 +
\omega_{2}\wedge e_2)+\cos \theta \cos \tilde\theta d\varphi
d\tilde\varphi ] \nn \\
\eea

\item R-R self-dual 5-form \be \hat{G}_5 = \cG_5 + * \cG_5 \quad
{\rm with} \quad \cG_5 = K(u) e_1\wedge e_2\wedge \omega_{1}\wedge
\omega_{2}\wedge \omega_{3} \ee where $\hat{G}_5 = G_5 + B_2\wedge
F_3$ with $G_5=dC_4$.

\end{itemize}

Integrating $\hat{G}_{5}=dC_4 + B_2\wedge F_3$ over a closed
`internal' 5-d section at fixed $u$ yields \be K(u) = N_3 +
2N_5[h_1(u) + b(u) h_2(u)]\ee \noindent that allows to eliminate
$K$ in terms of $b, h_1, h_2$ and the integers $N_3$ and $N_5$
(\ie number of D3- and D5-branes in the UV). The Bianchi identity
for $H_3$ yields \be dh_3 = {(e^{2g} + 2a^2 + e^{-2g} a^4 -
e^{-2g}) dh_1 + 2a(1 - e^{-2g} + a^2 e^{-2g}) dh_2 \over e^{2g} +
(1-a^2)^2e^{-2g} + 2a^2} \ee that allows one to eliminate $h_3$,
too.

The remaining scalar fields $\{p, {x}, g, a, b, \phi, h_1, h_2\}$
are governed by a 5-d effective Lagrangian with (almost) diagonal
metric $\cG_{ab}$ (only $h_1$ and $h_2$ mix with each other) and a
complicated potential that play no role in our analysis.

\subsection{Killing vectors}

For arbitrary choices of the functions ${x}, g, p, a, \phi, b,
h_1, h_2, (h_3, K)$ of the radial coordinate $u$, the metric and
$p$-forms are invariant under $SU(2)\times \widetilde{SU(2)}$
isometry generated by the six Killing vectors $\xi_a$
\bea \xi_{+} \equiv \xi_{1} &=& e^{i\varphi} \left(0,0,0,0,0,-1, 0, -i \cot\theta, 0, {i\csc\theta} \right) \nonumber \\
\xi_{-} \equiv \xi_{2} &=& e^{-i\varphi} \left(0,0,0,0,0,1, 0, -i \cot\theta, 0, {i\csc\theta} \right) \nonumber \\
\xi_{3} \equiv \xi_{3} &=& \left(0,0,0,0,0,0,0,1,0,0\right) \nonumber \\
\tilde\xi_{+} \equiv \xi_{4} &=& e^{i\tilde\varphi} \left(0,0,0,0,0,0,-1, 0, -i \cot\tilde\theta, {i\csc\tilde\theta} \right) \nonumber \\
\tilde\xi_{-} \equiv \xi_{5} &=& e^{-i\tilde\varphi} \left(0,0,0,0,0,0,1, 0, -i \cot\tilde\theta, {i\csc\tilde\theta} \right) \nonumber \\
\tilde\xi_{3} \equiv \xi_{6} &=& \left(0,0,0,0,0,0,0,0,1,0\right)
\eea Notice that $\xi_a$ have only components in the internal
directions, \ie $\xi^M = \delta^M_i \xi^i$ with $M=1,...,10$ and
$i=6,...,10$, and the contra-variant components displayed above
only depend on the internal `angular' variables, \ie $\de_\mu
\xi^M = 0$ with $\mu=1,...,5$. Although the metric does not mix
the angular variables with the non-compact variables, after
lowering the indices the components of the Killing vector acquire
a $u$ dependence due to warping. Clearly PT preserves Poincar\`e
symmetry in the `boundary' space-time directions, too.

It is easy to check that also the following two-forms
  \be
e_1 \wedge e_2 = -\sin\theta d\theta \wedge d\vf =
d\cos\theta\wedge d\vf \ee \be \omega_1 \wedge \omega_2 =
+\sin\tilde\theta d\tilde\theta \wedge d\tilde\vf = -
d\cos\tilde\theta \wedge d\tilde\vf \ee
 \be
\omega_1 \wedge e_1 + \omega_2 \wedge e_2= (\sin \psi \sin
\tilde\th d\tilde\varphi + \cos \psi d\tilde\th)\wedge d\theta +
\sin\theta (\sin \psi d\tilde\th - \cos \psi \sin \tilde\th
d\tilde\vf) \wedge d\vf \ee \be \omega_1 \wedge e_2 - \omega_2
\wedge e_1= - \sin\theta (\sin \psi \sin \tilde\th d\tilde\varphi
+ \cos \psi d\tilde\th)\wedge d\vf + (\sin \psi d\tilde\th - \cos
\psi \sin \tilde\th d\tilde\vf) \wedge d\theta\ee as well as the
one-form \be \tilde\omega_{3} = d\psi + \cos \tilde\th
d\tilde\varphi + \cos\th d\varphi \ee are $SU(2)\times
\widetilde{SU(2)}$ invariant, in the sense that
$\cL_\xi(...)=0$.\footnote{Lie derivatives act according to
$$\cL_v T_{M_1...M_p}{}^{N_1...N_q} = v^L\de_L
T_{M_1...M_p}{}^{N_1...N_q} + \sum_i
T_{M_1.L.M_p}{}^{N_1...N_q}\de_{M_i}  v^L - \sum_j
T_{M_1...M_p}{}^{N_1.L.N_q}\de_{L}  v^{N_j}$$.}

As a consequence, all background field-strengths are invariant \ie
\be \cL_{\xi_a} H_3 = 0 \quad , \quad \cL_{\xi_a} F_3 = 0 \quad ,
\quad \cL_{\xi_a} G_5 = 0 \ee Moreover, since \be \cL_{\xi_a} B_2
= 0 \ee one also has \be \cL_{\xi_a} \hat{F}_3 = 0 \quad , \quad
\cL_{\xi_a} \hat{G}_5 = 0 \ee while, in the chosen gauge, \be
\cL_{\xi_a} C_2 \neq 0 \ee By a change of gauge $\delta C_2 =
d\lambda^C_1$ we expect \be \cL_{\xi_a} C'_2 = 0 \ee Finally,
though rather obviously, $\cL_{\xi_a} \phi = 0$, $\cL_{\xi_a} \chi
= 0$.

While admitting $SU(2)\times \widetilde{SU(2)}$ isometry, the PT
ansatz generically `breaks' the abelian isometry associated to the
vector field $\hat\xi^M \de_M = \de/\de_\psi$. The latter may be
identified with the `anomalous' $U(1)$ R-symmetry of the dual $\cN
=1$ SYM theory on the boundary. In the bulk it is broken to
$Z_{2N}$ by the background 3-form and 5-form and then broken to
$Z_2$ by non-perturbative effects (string or D-brane instantons,
depending on the choice of wrapped branes). As discussed in
Section 2, the bulk counterpart of the anomalous divergence of the
R-symmetry current is a Higgs or rather St\"uckelberg mechanism
\cite{Klebanov:2002gr}, whereby a would-be massless vector field
eats an axion and becomes massive. This effect has been studied in
some details in \cite{Krasnitz:2000ir, Krasnitz:2002ct} in the
case of the KT background (a singular `relative' of KS solution),
confirming the expected value for the `mass' predicted by
\cite{Bianchi:2000sm}. For the case of MN solution, some
considerations about the required axion can be found in
\cite{Maldacena:2000yy, Gubser:2004qj}.

\subsection{Discrete symmetries and closed subsectors}

There are two $Z_2$ symmetries and their product that allow to
truncate Type IIB field equations in $D=10$ to closed sets of
fields mixing only with one another. The first is world-sheet
parity $\Omega$. The second is fermion parity in the L-moving
sector $(-)^{F_L}$, which is S-dual to $\Omega$ \ie $(-)^{F_L}=
S\Omega S^{-1}$, where S exchanges $F_3$ and $H_3$ and sends $\tau
= \chi + i e^{-\phi}$ to $-1/\tau$. The Einstein-frame metric and
the dilaton are invariant under both $\Omega$ and $(-)^{F_L}$,
while the action on the other bosonic fields is \bea
&&\Omega \qquad  (-)^{F_L} \nn \\
\chi \qquad &&- \qquad - \nn\\
B_2 \qquad &&- \qquad + \nn\\
C_2 \qquad &&+ \qquad - \nn\\
A_4 \qquad &&- \qquad - \nn \eea

Later on we will focus on the subsector invariant under $\Omega$.
For the PT ansatz,  this means \be h_1 = h_2 = 0 \rightarrow h_3=
K = 0 \ee MN solution for wrapped D5-branes
\cite{Maldacena:2000yy} belongs to this class, \ie it is invariant
under $\Omega$. Its dual wrapped NS5-brane solution belongs to the
class invariant under $(-)^{F_L}$. Standard $AdS_5 \times S^5$,
\ie near-horizon D3-branes, is invariant under $(-)^{F_L}\Omega$.
Finally KS and KT solutions (related to the conifold) do not
preserve any of the above discrete symmetries and are thus more
involved to study \cite{Berg:2005pd, Berg:2006xy}.

\section{Exact symmetries and Massless vectors}

In this Section, we would like to discuss the fate of the
${SU(2)}\times \widetilde{SU(2)}$ isometry, that should correspond
to the global `flavor' symmetry of the boundary theory, possibly
acting trivially on the lowest states relevant in the deep IR.
Even if from the vantage point of the holographic duality, the
presence of this isometry might be annoying, the analysis is quite
general and applies to any isometry in any flux compactification.

We will find that the Killing vectors generating
$\widetilde{SU(2)}$ are associated to truly massless vectors in
the bulk that correspond to an exact global `flavor' symmetry of
any solution based on PT ansatz. For the first $SU(2)$ factor the
situation is subtler, at least in the case of MN solution
\cite{Maldacena:2000yy}.

First of all notice that invariance of the metric under isometry
generated by a Killing vector $\xi^M$ reads \be \cL_\xi g_{MN}
\equiv \nabla_M \xi_N + \nabla_N \xi_M = 0 \ee that implies \be
\qquad \nabla_M \xi^M = 0 \ee as well as \be \nabla_M \xi_N =
{1\over 2}(\nabla_M \xi_N - \nabla_N \xi_M) \ee and \be \nabla_L
\nabla_M \xi_N = - R_{MNLK} \xi^K \ee

Invariance under diffeomorphisms suggests the existence of  a
trivial massless zero-mode for the metric fluctuations \be
\delta_{diff} g_{MN} = \nabla_{M}\beta_{N}+\nabla_{N}\beta_{M} \ee
Taking $\beta_{M}=\alpha(x)~\xi_{M}$ \be \delta_{diff} g_{MN} =
\xi_{N}\nabla_{M}\alpha+\xi_{M}\nabla_{N}\alpha \ee

It suggests an ansatz for the metric fluctuations of the form  \be
\delta_{phys} g_{MN} = -\xi_{N}A_{M} -\xi_{M}A_{N}\ee with $\xi^M
A_M = 0$ (\ie $\delta g^L{}_L = 0$) and $\cL_\xi A_M = 0$, since
$A_M = A_M(x)$ is to be independent of the five internal
coordinates the Killing vectors act on.  Gauge invariance under
$\delta A_{M}^{(0)} = -\nabla_{M}\alpha$ would then result from
general covariance and should imply massless-ness of the vector
field $A_M$. However, due to the presence of fluxes in the
background, the story is not so simple. The metric fluctuations
mix with $p$-form fluctuations, which we turn now our attention
onto.

Let us then consider the general case of an $n$-form $X_n$, whose
background $(n+1)$-form field strength $Y_{n+1} = dX_n$ is
invariant under some isometry generated by a Killing vector $\xi$
\be \cL_\xi Y_{n+1} = i_\xi dY_{n+1} + d(i_\xi Y_{n+1}) = 0 \ee
Thanks to Bianchi identity $dY_{n+1}=0$ one has (locally) \be
i_\xi Y_{n+1} = d Z^\xi_{n-1} \ee where $ Z^\xi_{n-1}$ is a
$(n-1)$-form defined up to an exact form $\delta{Z_{n-1}^\xi}=
dW_{n-2}$.

Under a diffeomorphism generated by $v^M = \alpha \xi^M$ \be \delta_{Diff} X_n = \alpha i_\xi Y_{n+1} + d(\alpha i_\xi X_n) = \alpha d
Z^\xi_{n-1}+ d(\alpha i_\xi X_n) = d\a\wedge Z^\xi_{n-1} + d[\alpha (i_\xi X_n - Z^\xi_{n-1})] \ee The last term can be cancelled by a gauge
transformation of the $n$-form $X_n$. This suggest that the correct ansatz for the `massless' vector $A_1$ associated to the coupled fluctuations
of the metric and $n$-form $X_n$ along $\xi$ be of the form \be \delta X_n = A_1 \wedge Z^\xi_{n-1} \ee In this way gauge invariance under $\delta
A_1 = d\alpha$ would not only be a consequence of general covariance but also of the $n$-form gauge invariance.  For the fluctuations of the
$(n+1)$-form field-strength $Y_{n+1}$ one then finds \be \delta Y_{n+1} = dA \wedge Z^\xi_{n-1}- A \wedge d Z^\xi_{n-1} = dA \wedge Z^\xi_{n-1} -
A_1 \wedge  i_\xi Y_{n+1} \ee

In principle the procedure applies to any background $n$-form in
PT or even more general flux vacua. The analysis can be performed
in quite general terms but it drastically simplifies in
backgrounds where $F_5=0, H_3=0, F_1=0$, thanks to invariance
under world-sheet parity $\Omega$, or else where $F_5=0, F_3=0,
F_1=0$, thanks to invariance under $(-)^{F_L}$. In both cases
mixing between $C_2$ and $B_2$ are excluded, and one can safely
set $A_4=0$ and even $\delta\phi=0$, as we will see.

Henceforth we will focus on the sub-sector invariant under
$\Omega$ \ie $C_2, g, \phi$ and set $B_2, A_4, \chi$ to zero both
in the background and in the fluctuations.

\subsection{$\Omega$ invariant Massless vectors (\ie $F_5 = 0$ and $H_3=0$)}

Taking into account that $\cL_\xi F_{3} =
d\left(i_{\xi}F_{3}\right) = 0 $, for any exact Killing vector,
one can locally write \be i_{\x}F_{3} = d\mu_1^\xi \ee that
suggests the following combined ansatze for the physical
fluctuations \be \delta g^{MN} = A^M \xi^N + A^N \xi^M \quad ,
\quad \delta C_{MN} = A_M \mu_N^{\xi}- A_N \mu_M^{\xi} \ee or,
equivalently, for the latter $\delta C_2 = A_1 \wedge \mu_N^{\xi}$
so that \be \delta F_{3} = d\delta C_{2} = dA\wedge \mu_1^{\xi}-A
\wedge i_{\xi}F_{3} \ee

Setting $\delta B_2 = 0, \delta \chi = 0, \delta A_4 =0$ as  well
as $\delta\phi = 0$, $\delta g_{\mu\nu} = 0, \delta g_{ij} =0$ one
has  $g^{MN} \delta g_{MN} = 0$, \ie $\delta \sqrt{||g||} = 0$.

Moreover, $\delta g_{MN} = -A_M \xi_N - A_N \xi_M $ so that \be
\delta g_{\mu i} = -A_\mu \xi_i =\delta g_{i\mu} \ee while \be
\delta C_{\mu i} = A_\mu \mu^\xi_i = - \delta C_{i\mu} \ee and
\bea
 \delta F_{ijk} &=& 0 \nonumber \\
 \delta F_{\mu\nu\rho} &=& 0 \nonumber \\
 \delta F_{\mu ij} &=& -A_{\mu} \left( \partial _{i} \mu^\xi_{j} - \partial _{j} \mu^\xi_{i}\right) \nonumber \\
 \delta F_{\mu \nu i} &=& \left(\partial_{\mu} A_{\nu} -
\partial_{\nu} A_{\mu} \right)\mu^\xi_{i} -
 \left(A_{\mu} \partial _{\nu} \mu^\xi_{i} - A_{\nu} \partial _{\mu} \mu^\xi_{i}\right)
 \eea

\subsubsection{Dilaton equation (Consistency check)}

Let us first check that it be consistent to set $\delta\phi =0$.
Using the 3-form ansatz, one finds
\bea \delta F^2 &=& \delta F_{LMN}~F^{LMN} + 3~F_{LMN} \delta g^{LP}F_{P}{}^{MN} \nonumber \\
&=& F^{ijk} \delta F_{ijk} + 3 F^{ujk}~\delta F_{ujk} + 6 F_{ujk} \delta g^{ui} F_{i}{}^{jk}
\nn \\
&=& -3 A_{u}~F^{ujk}\left( \partial _{j} \mu^\xi_{k} - \partial
_{k} \mu^\xi_{j}\right) + 6 \delta g^{ui}~F_{ujk} F_{i}{}^{jk}
\eea

For a solution of the PT kind one has $F_{ujk} F_{i}{}^{jk} = 0$,
$i = {6,..., 10}$. Also one obtains $F^{ujk}\left(
\partial _{j} \mu^\xi_{k} - \partial _{k} \mu^\xi_{j}\right) = 0$
for each one of the six Killing vectors $\xi_a$. Therefore $\delta
F^2 = 0$, consistently with the ansatz $\delta \f = 0$.

\subsubsection{3-form equation}
Let us focus on the 3-form equation \bea \delta \left(\nabla_{M}
\left(\sqrt{g}e^{\phi}F^{MNP}\right)\right) &=& {1\over
\sqrt{g}}\partial_{M}
\left(\sqrt{g} e^{\phi}\left[\delta g^{ML} F_{L}{}^{NP} + \delta g^{NL} F^{M}{}_{L}{}^{P} \nonumber\right.\right.\\
&&\left.\left.+ \delta g^{PL} F^{MN}{}_{L} + g^{ML}g^{NK}g^{PQ}\delta F_{LKQ}
 \right]\right)\eea

Decomposing into space-time ($\nu=1,...,5$) and internal indices ($j=6,...,10$) one has

\begin{itemize}

\item Equations $N=\nu$, $P=\rho$ \bea \delta \left(\nabla_{M}
\left(\sqrt{g}e^{\phi}F^{M\nu\rho}\right)\right) &\equiv& {1\over
\sqrt{g}}\partial_{M} \left(\sqrt{g} e^{\phi}\left[\delta g^{ML}
F_{L}{}^{\nu \rho} + \delta
g^{\nu L} F^{M}{}_{L}{}^{\rho} \nonumber \right.\right.\\
&&\left.\left. + \delta g^{\rho L} F^{M\nu}{}_{L} + g^{ML}g^{\nu
K}g^{\rho Q}\delta F_{LKQ} \right]\right) = 0 \eea Keeping only
non-vanishing components yields \be {1\over \sqrt{g}}\partial_{M}
\left(\sqrt{g} e^{\phi}\left[\delta g^{\nu l} F^{M}{}_{l}{}^{\rho}
+ \delta g^{\rho L} F^{M\nu}{}_{L} + g^{ML}g^{\nu k}g^{\rho
\mu}\delta F_{Lk\mu} \right]\right) = 0 \ee Plugging the anstaze
one eventually finds  \be e^{\phi}
\partial_{i} \left\{ \sqrt{\hat{g}}\left[A^{\nu} \xi^{l}
F^{i}{}_{l}{}^{u}\delta^{\rho}_{u} + A^{\rho} \xi^{l}
\delta^{\nu}_{u} F^{i u}{}_{l} + g^{il}g^{\nu \lambda}g^{\rho
\sigma}\left(\mu^\xi_{l}
f_{\lambda\sigma}-A_{\lambda}\partial_{\sigma}\mu^\xi_{l} +
A_{\sigma}\partial _{\lambda} \mu^\xi_{l}\right)\right] \right\} =
0 \ee where $f_{MN} =
\partial_{M}A_{N}-\partial_{N}A_{M}$ and $\sqrt{\hat{g}}$ denote
the dependence of $\sqrt{{g}}$ on the internal coordinates.
Finally, using $i_\xi F_3 = d\mu_\xi$ one arrives at the following
constraint for $\mu_\xi$ \be
0=\partial_{i}\left(\sqrt{\hat{g}}\mu^{i}\right)f^{\nu\rho}
\Rightarrow \nabla_{M}\mu^{M} = 0 \ee One can check that this is
satisfied for all $\mu_\xi$ in any background of the PT kind.
Anyway, we expect that it should always be possible to satisfy the
constraint by adding to $\mu^\xi$  an exact form $d\eta^\xi$,
where $\eta^\xi$ is an appropriate function.

\item Equations for $N=i$, $P=j$ \bea \delta \left(\nabla_{M}
\left(\sqrt{g} e^{\phi}F^{M i j}\right)\right) &=& {1\over
\sqrt{g}}\partial_{M} \left(\sqrt{g} e^{\phi}\left[\delta g^{ML}
F_{L}{}^{ij} + \delta g^{iL}
F^{M}{}_{L}{}^{j} \nonumber \right.\right.\\
 &&\left.\left.+ \delta g^{jL} F^{Mi}{}_{L} + g^{ML}g^{iK}g^{jQ}\delta F_{LKQ}
 \right]\right) = 0 \eea
Keeping only non-vanishing components yields \be {1\over
\sqrt{g}}\partial_{M} \left(\sqrt{g} e^{\phi}\left[\delta g^{ML}
F_{L}{}^{ij} + \delta g^{iL} F^{M}{}_{L}{}^j - \delta g^{jL}
F^{M}{}_{L}{}^i + g^{ML}g^{i k}g^{j l}\delta F_{Lkl}
\right]\right) = 0 \ee

Plugging the anstaze one eventually finds \bea
&&{e^{\phi}\over\sqrt{\hat{g}}} A^{u}\partial_{m} \left\{
\sqrt{\hat{g}}\left[\xi^m F_{u}{}^{ij}+ \xi^{i} F^{m}_u{}^{j}
-\xi^{j} F^{m}_u{}^{i}\right]\right\} \nn \\ &&\quad +
{1\over\sqrt{{g}}}
\partial_{\mu} \left\{ e^{\phi} \sqrt{{g}}\left[A^{\mu}\xi^k
F_{k}{}^{ij}- A^{\mu} (\nabla^i\mu^j_\xi -  \nabla^i\mu^j_\xi)
\right] \right\} = 0 \eea that is satisfied after using $\cL_\xi
F_3 =0$ \ie $i_\xi F_3 = d\mu_\xi$, $\nabla_M \xi^M = 0$ and the
background 3-form equation $\partial_{m} (
\sqrt{\hat{g}}F_{u}{}^{mj}) = 0$.

\item Equations for $N=\nu$, $P=l$
\bea
\delta \left(\nabla_{M} \left(\sqrt{g} e^{\phi}F^{M\nu
l}\right)\right) &=& {1\over \sqrt{g}}\partial_{M} \left(\sqrt{g}
e^{\phi}\left[\delta g^{ML} F_{L}{}^{\nu
l} + \delta g^{\nu L} F^{M}{}_{L}{}^{l} \nonumber\right.\right.\\
&&\left.\left.+ \delta g^{lL} F^{M\nu}{}_{L} + g^{ML}g^{\nu K}g^{lQ}\delta
F_{LKQ}
 \right]\right)= 0 \eea
 Keeping only non-zero components yields
\bea 0&=& {1\over \sqrt{g}}\partial_{\mu}\left(\sqrt{g}
e^{\phi}\left[\delta g^{\mu k} F_{k}{}^{\nu l} + \delta g^{\nu k}
F^{\mu}{}_{k}{}^{l} + g^{\mu
\rho} g^{\nu \lambda}g^{lj}\delta F_{\rho\lambda j} \right]\right)\nonumber\\
&+& e^{\phi} \partial_{i} \left(\delta g^{\nu k} F^{i}{}_{k}{}^{l}
+ g^{ik}g^{\nu \mu}g^{lj}\delta F_{k\mu j}\right).
 \eea
Plugging in the ansatz for the fluctuations yields \bea && {1\over
\sqrt{g}}\partial_{\mu} \left(\sqrt{g} e^{\phi}\left[A^{\mu}
\xi^{k} F_{k}{}^{\nu l} + A^{\nu} \xi^{k} F^{\mu}{}_{k}{}^{l} +
g^{\mu\rho} g^{\nu \lambda}g^{lj}\left(f_{\rho\lambda}\mu_{j} -
A_{\rho}\xi^{n}F_{n\lambda
j} \nonumber\right.\right.\right. \\
&&\left.\left.\left.\quad + A_{\lambda}\xi^{n}F_{n\rho
j}\right)\right]\right) + e^{\phi} \partial_{i} \left(A^{\nu}
\xi^{k}F^{i}{}_{k}{}^{l} + g^{ik}g^{\nu
\mu}g^{lj}\left(-A_{\mu}\xi^{n}F_{njk}\right)\right)= 0 \eea After
various cancellations, one finally arrives at the Dynamical
Equation for the vector fields \be
\partial_{\mu}\left(\sqrt{g}e^{\phi} g^{\mu\rho}
g^{\nu\lambda}f_{\rho\lambda}\mu_{\xi}^{l}\right) = 0 \ee
\end{itemize}
When $\mu^i_\xi = e^{-\phi/2} \xi^i$, this further simplifies into
\be {\xi}^{i}\partial_{\mu}\left(\sqrt{g}e^{\phi/2} g^{\mu\rho}
g^{\nu\lambda}f_{\rho\lambda}\right) = 0 \ee which neatly displays
the correspondence between bulk 5-d massless vector fields and
exact Killing vectors.

\subsubsection{Einstein equations}

It is straightforward but very laborious to show that Einstein
equations  lead to the same results, \ie the very same dynamical
equation for $A_\mu$. For simplicity we will restrict our
attention on the case in which $\mu_M^\xi = e^{-\phi/2} \xi_M$ \ie
$i_{\xi}F_{3}=d(e^{-{\phi\over2}}\lambda^\xi)$ where $\lambda^\xi
= g_{MN}\xi^Mdx^N$

Let us start with the source term \bea &&\delta S_{MN} =
{1\over4}e^{\phi}\left(\delta F_{MPQ} F_{N}{}^{PQ} + \delta
F_{NPQ} F_{M}{}^{PQ} + 2\delta g^{PL} F_{MPQ}
F_{NL}{}^{Q}\right)-{1\over48}e^{\phi}\delta g_{MN}F^{2}\nonumber \\
&&\quad =
{1\over4}\left[e^{\phi\over2}F_{N}{}^{PQ}\left(f_{[MP}\xi_{Q]}-e^{\phi\over2}A_{[M}F_{
PQ]L}\xi^{L}\right) +
e^{\phi}\left(A^{P}\xi^{L}+A^{L}\xi^{P}\right)F_{MPQ}F_{NL}{}^{Q} \right. \nn \\
&& \quad \left. + (M\leftrightarrow N)\right]
-{1\over48}e^{\phi}\delta g_{MN}F^{2}\eea

Defining \be h_{LM} = \nabla_{L}\xi_{M}-\nabla_{M}\xi_{L}=
{1\over2}\left(\partial_{L}\phi\xi_{M}-\partial_M{\phi}\xi_{L}\right)
+ e^{\phi}\left(e^{-{\phi\over2}}\xi^{P}F_{PLM}\right) \ee the
first order fluctuation of the Ricci tensor reads \bea &&\delta
R_{MN} = -{1\over2}\left(\nabla^{2}\delta g_{MN} -
\nabla^{L}\nabla_{M} \delta g_{NL} - \nabla^{L}\nabla_{N}\delta
g_{ML} +
\nabla_{M}\nabla_{N}\delta g^{L}_{L}\right)\nonumber \\
&& = -{1\over2}\nabla^{L}\left[\xi_{N}f_{ML} + \xi_{M}f_{NL} +
\xi_{L}\left(\nabla_{M}A_{N} + \nabla_{N}A_{M}\right) +
A_{M}h_{NL} +
A_{N}h_{ML}\right]\nonumber \\
&& =
{1\over2}e^{\phi\over2}\left[f_{LM}\nabla^{L}\left(e^{-{\phi\over2}}\xi_{N}\right)
+ f_{LN}\nabla^{L}\left(e^{-{\phi\over2}}\xi_{M}\right)\right] +
{1\over2}e^{-{\phi\over2}}\left[\xi_{N}\nabla^{L}\left(e^{\phi\over2}f_{LM}\right)
\nonumber \right.\\
&&\left.+
\xi_{M}\nabla^{L}\left(e^{\phi\over2}f_{LN}\right)\right] -
{1\over2}\xi^{L}\nabla_{L}\left(\nabla_{M}A_{N} +
\nabla_{N}A_{M}\right) + {1\over2}\nabla^{L}\left(A_{M}h_{LN} +
A_{N}h_{LM}\right)\eea Moreover one finds \bea
&&{1\over2}\nabla^{L}\left(A_{M}h_{LN} + A_{N}h_{LM}\right)
={1\over2}\nabla^{L}\left(A_{M}e^{\phi\over2}\xi^{P}F_{PLN} +
A_{N}e^{\phi\over2}\xi^{P}F_{PLM}\right)\nonumber\\
&+& {1\over4}\nabla^{L}\left(A_{M}\xi_{N}\partial_{L}\phi - A_{M}\xi_{L}\partial_{N}\phi +
A_{N}\xi_{M}\partial_{L}\phi-A_{N}\xi_{L}\partial_{M}\phi\right)\nonumber \\
&=&{1\over2}e^{\phi\over2}\xi^{P}F_{PLN}\nabla^{L}A_{M}-{1\over4}e^{\phi}\xi^{k}F_{k}{}^{LP}F_{LPN}A_{M}-{1\over48}e^{\phi}\delta
g_{MN}F^{2}\nonumber\\
&&+{1\over4}\partial_{L}\phi\nabla^{L}\left(A_{M}\xi_{N}+A_{N}\xi_{M}\right)-{1\over4}\xi^{L}\nabla_{L}
\left(A_{M}\partial_{N}\phi+A_{N}\partial_{M}\phi\right) \eea and
also \be -{1\over4}e^{\phi}\xi^{k}F_{k}{}^{LP}F_{LPN}A_{M} =
-{1\over4}e^{\phi}F_{N}{}^{PQ}\xi^{L}A_{[M}F_{L PQ]} + {1\over4}
e^{\phi}\left(A^{P}\xi^{L}+A^{L}\xi^{P}\right)F_{MPQ}F_{NL}{}^{Q}
\ee Simplifying terms in the left and right hand side yields \bea
\delta' S_{MN} &=&
{1\over4}e^{\phi\over2}F_{N}{}^{PQ}f_{[MP}\xi_{Q]} +
(M\leftrightarrow N) \eea and \bea \delta' R_{MN} &=&
{1\over2}e^{\phi\over2}f_{LM}\nabla^{L}\left(e^{-{\phi\over2}}\xi_{N}\right)
+
{1\over2}e^{-{\phi\over2}}\xi_{N}\nabla^{L}\left(e^{\phi\over2}f_{LM}\right)-{1\over2}\xi^{L}\nabla_{L}\nabla_{M}A_{N}\\
&&+ {1\over2}e^{\phi\over2} \xi^{P}F_{PLN}\nabla^{L}A_{M}
+{1\over4}\partial_{L}\phi\nabla^{L}\left(A_{M}\xi_{N}\right)-{1\over4}\xi^{L}\nabla_{L}\left(A_{M}\partial_{N}\phi\right)+ (M\leftrightarrow
N)\nonumber\eea

In particular for $M=\mu$ and $N=i$, the remaining terms in the source fluctuations are \be \delta S_{\mu i} = {1\over2}
e^{-{\phi\over2}}\left({1\over2}\phi^{'}\xi_{i}-\xi_{i}^{'}\right)\left(\nabla_{\mu}A_{u}-\nabla_{u}A_{\mu}\right)\ee while the remaining terms in
the Ricci fluctuations read \be {1\over2}e^{\phi\over2}f_{LM}\nabla^{L}\left(e^{-{\phi\over2}}\xi_{N}\right) + (M\leftrightarrow N) =
{1\over4}e^{-{\phi\over2}}\left(\nabla_{u}A_{\mu}- \nabla_{\mu}A_{u}\right)\left(\xi_{i}^{'}-\phi^{'}\xi_{i}\right) \quad , \ee \be
-{1\over2}\xi^{L}\nabla_{L}\nabla_{M}A_{N} + (M\leftrightarrow N)=
-{1\over4}e^{-{\phi\over2}}\xi_{i}^{'}\left(\nabla_{\mu}A_{u}+\nabla_{u}A_{\mu}\right) + {1\over4}e^{-{\phi\over2}} \delta_{\mu}^{u}
\xi_{j}^{'}g^{jk}A_{u}\partial_{u}g_{ki} \quad ,\ee
\bea &&{1\over2}e^{\phi\over2} \xi^{P}F_{PLN}\nabla^{L}A_{M} + (M\leftrightarrow N) =\nonumber \\
&&e^{-{\phi\over2}}\left\{-{1\over4}\phi^{'}\xi_{i}\nabla_{u}A_{\mu} + {1\over2}\xi_{i}^{'}\nabla_{u}A_{\mu}
-\delta_{\mu}^{u}\left(-{1\over8}\phi^{'}A_{u}\xi_{i}^{'} + {1\over4} \xi_{j}^{'}g^{jk}A_{u}\partial_{u}g_{ki}\right) \right\} \quad , \eea \be
{1\over4}\partial_{L}\phi\nabla^{L}\left(A_{M}\xi_{N}\right) + (M\leftrightarrow N) =
e^{-{\phi\over2}}\left({1\over4}\phi^{'}\xi_{i}\nabla_{u}A_{\mu}+ {1\over8}\xi_{i}^{'}\phi^{'}A_{\mu}\right)\ee and finally \be
-{1\over4}\xi^{L}\nabla_{L}\left(A_{M}\partial_{N}\phi\right)+ (M\leftrightarrow N) =
e^{-{\phi\over2}}\left(-{1\over8}\xi_{i}^{'}A_{\mu}\phi^{'}-{1\over8}\xi_{i}^{'}A_{u}\phi^{'}\delta_{\mu}^{u}\right)\ee

After a number of cancellation one eventually finds \be\delta
S_{\mu i}= \delta R_{\mu i}\ee when the 3-form equations are
satisfied, showing that the fluctuations ansatz then satisfies
Einstein equations, too.

It is easy to check that the other components $(i,j)$ and
$(\mu,\nu)$ are satisfied as well. For brevity, we refrain to
present the details here.

\subsection{One-forms associated to Killing vectors}

For completeness and later use, let us display in the following
the expressions of the one-forms associated to the Killing vectors
in a generic PT background. Setting\footnote{The constant factor
$N_{5}^{-1}$ is inserted for later notational convenience.}
$\lambda_{\xi} = N_{5}^{-1} g_{ij} \xi^{i}dx^{j}$, one has \bea
\lambda_1 &=& -\frac{1}{4} e^{i \varphi -g +x } \left(a^2+4 e^{2 g
}\right)d\theta + \frac{1}{4} e^{i \varphi -g +x } a (\cos \psi+i
\cos
   \th  \sin \psi)d\tilde\theta \nonumber \\
   &-& \frac{1}{4} i e^{i \varphi -g -6 p -x } \left(e^{6
   p +2 x } a ^2-e^{g }+4 e^{2 (g +3 p +x )}\right) \cos \th  \sin \th d\varphi
   \nonumber \\
   &+& \frac{1}{4} e^{i \varphi -g -6 p -x } \left[i e^{g } \cos\tilde\th\sin \th
   + e^{6 p+2 x} a (\sin \psi-i \cos \psi
   \cos \th ) \sin \tilde\th \right]d\tilde\varphi\nonumber \\
   &+& \frac{1}{4} i e^{i \varphi -6 p-x} \sin \th d\psi\eea
\bea \lambda_2 &=& \frac{1}{4} e^{-i \varphi -g+x} \left(a^2+4
e^{2
   g}\right)d\theta -\frac{1}{4} e^{-i \varphi -g+x} a (\cos \psi-i
\cos\th  \sin \psi) d\tilde\theta \nonumber \\
   &-& \frac{1}{4} i e^{-i \varphi -g-6 p-x} \left[e^{6
   p+2 x} a^2-e^{g}+4 e^{2 (g+3 p+x)}\right] \cos \th  \sin
   \th d\varphi\nonumber \\
   &+& \frac{1}{4} i e^{-i \varphi -g-6 p-x} \left[e^{g} \cos
   \tilde\th  \sin \th -e^{6 p+2 x} a (\cos \psi \cos \th -i \sin \psi) \sin \tilde\th
   \right]d\tilde\varphi\nonumber \\
   &+& \frac{1}{4} i e^{-i\varphi -6 p-x} \sin \th d\psi\eea
\bea \lambda_3 &=& -\frac{1}{4} e^{x-g} a \sin \psi \sin
   \th d\tilde\theta + \frac{1}{4} \left[e^{-6 p-x} \cos ^2\th +e^{x-g}
   \left(a^2+4 e^{2 g}\right) \sin ^2\th \right] d\varphi \nonumber \\
   &+& \frac{1}{4} \left(e^{-6p-x}
\cos \th  \cos \tilde\th +e^{x-g} a \cos \psi \sin \th  \sin
\tilde\th \right)d\tilde\varphi+ \frac{1}{4} e^{-6 p-x} \cos \th
d\psi \eea
 \bea
\lambda_4 &=& \frac{1}{4} e^{i \tilde\varphi -g+x} a (\cos \psi+i
\cos
   \tilde\th  \sin \psi)d\theta -\frac{1}{4} e^{i \tilde\varphi -g+x} d\tilde\theta
   \nonumber \\
&+& \frac{1}{4} e^{i\tilde\varphi -g-6 p-x} \left[e^{6 p+2 x} a
(\sin \psi-i \cos \psi \cos
   \tilde\th ) \sin \th +i e^{g} \cos \th  \sin \tilde\th \right] d\varphi\nonumber \\
&-& \frac{1}{8} i e^{i\tilde\varphi -g-6 p-x}
   \left(-e^{g}+e^{6 p+2 x}\right) \sin (2 \tilde\varphi)d\tilde\varphi +
   \frac{1}{4} i e^{i\tilde\varphi -6 p-x} \sin
\tilde\th d\psi \eea
 \bea \lambda_5 &=&
-\frac{1}{4} e^{-i \tilde\varphi -g+x} a (\cos \psi-i \cos
   \tilde\th  \sin \psi)d\theta + \frac{1}{4} e^{-i \tilde\varphi -g+x}
d\tilde\theta\nonumber \\
&-& \frac{1}{4} i e^{-i\tilde\varphi -g-6 p-x} \left[(e^{6 p+2 x}
a (\cos \psi \cos
   \tilde\th -i \sin \psi) \sin \th -e^{g} \cos \th  \sin \tilde\th \right]d\varphi\nonumber \\
&+& \frac{1}{4} i e^{-i\tilde\varphi -g-6 p-x} \left(e^{g}-e^{6
   p+2 x}\right) \cos \tilde\th  \sin \tilde\th d\tilde\varphi +
   \frac{1}{4} i e^{-i\tilde\varphi -6 p-x} \sin
\tilde\th d\psi \eea \bea \lambda_6 &=& -\frac{1}{4} e^{x-g} a
\sin \psi \sin
   \tilde\th d\theta
   + \frac{1}{4} \left[e^{-6 p-x} \cos \th  \cos
   \tilde\th +e^{x-g} a \cos \psi \sin \th  \sin
   \tilde\th \right]d\varphi \nonumber \\
   &+& \frac{1}{4} \left[e^{-6 p-x} \cos^2\tilde\th
   + e^{x-g} \sin ^2\tilde\th \right]d\tilde\varphi+  \frac{1}{4} e^{-6 p-x}\cos \tilde\th
   d\psi\eea

Then using $N_{5}^{-1} i_\xi F_3 = d\mu^\xi$ one finds \bea
\mu_{1} &=& \frac{1}{4} \left[e^{i \varphi} f- e^{i \varphi}
(f-1)\right]d\theta  -\frac{1}{4}
e^{i\varphi} b (\cos \psi + i \cos \th \sin \psi) d\tilde\theta \nonumber \\
&&+ \frac{1}{8}\left[ i e^{i\varphi} (f-1) \sin (2 \th)-2 i e^{i \varphi} \cos \th(f-1)
\sin \th\right] d\varphi \nonumber\\
&&-\frac{1}{4} e^{i \varphi} \left[i \cos \tilde\th \sin \th +
b (\sin \psi -i \cos \psi \cos \th) \sin \tilde\th \right]d\tilde\varphi \nonumber \\
&&-\frac{1}{4} i e^{i \varphi} \sin \th d\psi\eea \bea \mu_{2} &=&
\frac{1}{4} \left[e^{-i \varphi} (f-1)- e^{-i \varphi} f \right]
d\theta + \frac{1}{4}
   e^{-i \varphi} b  (\cos \psi -i \cos \th  \sin \psi )d\tilde\theta \nonumber \\
&&+ \frac{1}{8} \left[i e^{-i\varphi} (f -1) \sin (2 \th)-2 i
e^{-i \varphi} \cos \th
   (f -1) \sin \th \right] d\varphi\nonumber\\
   &&-\frac{1}{4} i e^{-i \varphi} \left[\cos \tilde\th  \sin
   \th + b  (i \sin \psi -\cos \psi  \cos \th ) \sin
   \tilde\th \right]d\tilde\varphi\nonumber \\
&&-\frac{1}{4} i e^{-i \varphi} \sin \th d\psi\eea \bea \mu_{3}
&=&\frac{1}{4} b  \sin \psi  \sin \th d\tilde\theta + \frac{1}{4}
\left[(f -1) \sin ^2\th
+ \left(-\cos ^2\th -f\sin^2\th \right)\right]d\varphi\nonumber\\
&&+ \frac{1}{4} (-\cos \th  \cos \tilde\th - b \cos \psi  \sin\th
\sin \tilde\th )d\tilde\varphi -\frac{\cos \th }{4}d\psi\eea \bea
\mu_{4} &=&\frac{1}{4} e^{i \tilde\varphi} b (\cos \psi +i \cos
\tilde\th  \sin \psi)d\theta -\frac{e^{i
\tilde\varphi}}{4}d\tilde\theta
\nonumber \\
&&\frac{1}{4} e^{i \tilde\varphi}\left[ b (\sin \psi -i \cos \psi
\cos
   \tilde\th ) \sin \th +i \cos \th  \sin \tilde\th \right]d\varphi +
   \frac{1}{4} i e^{i\tilde\varphi} \sin \tilde\th d\psi \eea
\bea \mu_{5}&=&-\frac{1}{4} e^{-i \tilde\varphi} b (\cos \psi -i
\cos \tilde\th  \sin \psi )d\theta
+ \frac{e^{-i \tilde\varphi}}{4}d\tilde\theta\\
   &&-\frac{1}{4} i e^{-i \tilde\varphi} \left[ b  (\cos \psi  \cos
   \tilde\th -i \sin \psi ) \sin \th -\cos \th  \sin\tilde\th
   \right]d\varphi
   + \frac{1}{4} i e^{-i \tilde\varphi} \sin \tilde\th d\psi \nonumber \eea
\bea \mu_{6}&=&-\frac{1}{4} b  \sin \psi  \sin\tilde\th d\theta +
\frac{1}{4} (\cos \th
   \cos \tilde\th + b  \cos \psi  \sin \th  \sin\tilde\th )d\varphi\nonumber \\
&&+ \frac{1}{4}d\tilde\varphi + \frac{1}{4}\cos\tilde\th d\psi\eea
which are valid in any PT background. We also defined $f = 4
e^{2g} + a^{2}$.

For MN background \cite{Maldacena:2000yy} a compact form for
$\mu^{\xi}$ in terms of the (rescaled) KV obtains
\bea \mu_{(1)i}&=&-e^{-{\phi\over2}}\xi_{(1)i} - {1\over4}\left(f-1\right)e^{i\varphi}\left(\delta_{i}^{6} + i \cos \theta \sin \theta \delta_{i}^{8}\right) \nonumber \\
\mu_{(2)i}&=&-e^{-{\phi\over2}}\xi_{(2)i} + {1\over4}\left(f-1\right)e^{-i\varphi}\left(\delta_{i}^{6} - i \cos \theta \sin \theta \delta_{i}^{8}\right)\nonumber \\
\mu_{(3)i}&=&-e^{-{\phi\over2}}\xi_{(3)i} + {1\over4}\left(f-1\right)\sin^2 \theta \delta_{i}^{8}\nonumber \\
\mu_{(4)i}&=&e^{-{\phi\over2}}\xi_{(4)i}\nonumber \\
\mu_{(5)i}&=&e^{-{\phi\over2}}\xi_{(5)i}\nonumber \\
\mu_{(6)i}&=&e^{-{\phi\over2}}\xi_{(6)i} \eea

\subsection{Scalar products and gauge kinetic functions}

The scalar products of the Killing Vectors are diagonal as a
result of the $SU(2)\times\widetilde{SU(2)}$ symmetry. For $SU(2)$
one finds  \be \int \xi^{* i}_{a} g_{ij} \xi^{j}_{b} d^5\Omega =
\frac{2}{3} N_{5}^{7\over2} \pi ^3 \delta_{ab} \kappa_a e^{-g -9 p
+\frac{x }{2}} \left[2 e^{6 p +2 x }(f-1)+ \left(e^{g} + 2 e^{6 p
+2 x}\right)\right]\ee with $\kappa_1 =\kappa_2 = 2$ and $\kappa_3
= 1$ , while for $\widetilde{SU(2)}$ one finds \be \int \xi^{*
i}_{a} g_{ij} \xi^{j}_{b} d^5\Omega = \frac{2}{3} N_{5}^{7\over2}
\pi ^3 \delta_{ab} \kappa_a e^{-g -9 p +\frac{x }{2}} \left(e^{g
}+2 e^{6 p +2 x }\right) \ee with $\kappa_4 =\kappa_5 = 2$ and
$\kappa_6 = 1$.

Similarly for the one-forms $\mu^\xi_a$, the scalar products are
diagonal. For $SU(2)$ one gets \be \int \mu^{*
i}_{a}g_{ij}\mu^{j}_{b} d^5\Omega = \frac{1}{3} e^{-g -3 p
+\frac{x }{2}} \delta_{ab} \kappa_a N_{5}^{7\over2} \pi ^3
\left[\left(f-1\right)\left(a^2-1\right)+2e^g\left(2 e^{g}+e^{6p +
2x}\right)\right] \ee with $\kappa_1 =\kappa_2 = 2$ and $\kappa_3
= 1$, while for $\widetilde{SU(2)}$ one gets \be \int \mu^{*
i}_{a}g_{ij}\mu^{j}_{b} d^5\Omega = \frac{2}{3}
N_{5}^{7\over2}\pi^3 e^{\frac{x }{2}-3 p } \delta_{ab} \kappa_a
\left(2 e^{g }+e^{6 p +2 x }\right) \ee with $\kappa_4 =\kappa_5 =
2$ and $\kappa_6 = 1$, exposing the $SU(2)\times\widetilde{SU(2)}$
symmetry.

From the above scalar products one can read the gauge kinetic
functions for the massless vectors of
$SU(2)\times\widetilde{SU(2)}$ in 5-d. For $SU(2)$ one finds  \bea
\cK_{SU(2)} = {1\over 2} \cV^{-{1\over 3}}(||\xi||^2 + e^\phi||\mu||^2)
&=& {N_5^{8\over3}\over12}\pi^2 \delta_{ab}\kappa_a e^{-2p-g}\left\{(f-1)
\left[4e^{2x}+e^{\phi}\left(a^2-1\right)\right]\nonumber\right.\\
&&\left.+\left[2e^{g-6p}+4e^{2x}+2e^{g+\phi}
\left(2e^{g}+e^{6p+2x}\right)\right]\right\}\eea
where the first term is a contribution from the Einstein-Hilbert
term, while the second comes from the $F_3^2$ in the Type IIB
action\footnote{Which is well defined for $A_4=0$.}. For
$\widetilde{SU(2)}$, $\mu = e^{-\phi/2} \xi$ and one simply finds
\be \cK_{\widetilde{SU(2)}} = \cV^{-{1\over 3}}||\xi||^2 =
\frac{N_5^{8\over3}\pi^2}{3}\delta_{ab}\kappa_a
e^{-g-8p}\left(e^g+2e^{6p+2x}\right)\ee The internal volume factor
$\cV^{-{1\over 3}}$ arises from the Weyl scaling of the 10-d
metric with pure space-time components so as to have canonical E-H
term in 5-d, \ie $g^{(10)}_{\mu\nu} = \cV^{-{2\over 3}}(u)
g^{(5)}_{\mu\nu}$.

\section{Massless Vectors in MN background}

We will now explicitly apply the above analysis to the case of MN
solution for wrapped D5-branes \cite{Maldacena:2000yy}. For
simplicity we will focus on the $\widetilde{SU(2)}$ factor, for
which $\mu^\xi = e^{-\phi/2} \xi$.

\subsection{MN Solution}

In MN solution for wrapped D5-branes \cite{Maldacena:2000yy} one
has $h_1 = h_2 = 0$ (no D3-branes $F_5=0$, no NS5-branes $H_3 =0$,
$\chi = 0$) and $b=a$. Denoting the radial variable by $u$, the
metric reads
 \be
 ds^{2} = e^{\phi\over 2} \left[dx^2 + N_{5}\left\{du^2 + e^{2g}\left(e_{1}{}^{2}+e_{2}{}^{2}\right) + {1\over 4}\left(\tilde\w_{1}{}^{2} +
 \tilde\w_{2}{}^{2} + \tilde\w_{3}{}^2 \right)\right\} \right]
 \ee
where
\begin{eqnarray}
 &&e^{-2\f} = {2 e^{g} \over \sinh 2u} \quad , \quad
 e^{2g} = u ~ \coth 2u - {1 \over 4} (1 + a^2) \nonumber \\
 && a = {2 u \over \sinh 2u} \quad , \quad
 f =  4 e^{2g}+a^2 \nonumber \\
  \end{eqnarray}
The RR 3-form flux is given by \bea F_3 &=& {N_5\over 4}
\left\{\tilde\omega_{3}\wedge [(\omega_{1}\wedge \omega_{2} +
e_1\wedge e_2) - a(u)
(\omega_{1}\wedge e_2 - \omega_{2}\wedge e_1)] \nonumber \right.\\
&&\left.+ a'(u) du \wedge\left(\omega_{1}\wedge e_1 +
\omega_{2}\wedge e_2\right)\right\} \eea

The asymptotic behavior of the radial functions in the UV
($u\rightarrow 0$) and IR ($u\rightarrow \infty $) are found to be
\be a(u \rightarrow 0) \rightarrow 1 - {2\over 3} u^2 \quad ,
\quad a(u \rightarrow \infty) \rightarrow 0\ee \be e^{2g}(u
\rightarrow 0) \rightarrow u^{2} \quad , \quad e^{2g}(u
\rightarrow \infty) \rightarrow u\ee \be f(u \rightarrow 0)
\rightarrow 1 + {8\over 3} u^2 - \frac{32}{45}u^4\quad , \quad f(u
\rightarrow \infty) \rightarrow 4u\ee

For later use, notice that \bea && g_{\mu\nu} = e^{\phi \over 2} \delta_{\mu\nu} \quad , \quad g_{ij} = e^{\phi \over 2} \hat{g_{ij}} \quad ,
\quad det ~ g_{MN} = {N_{5}^{6}\over 64} e^{5\f}e^{4g}\sin^{2}\th ~ \sin^{2}\tilde\th \nn
 \eea

\subsection{Spectrum of massless vector harmonics}

As we have seen, in order to find the spectrum of bound-states
that are holographically dual to the massless bulk vectors
associated to the three Killing vectors of $\widetilde{SU(2)}$,
one should solve \be {1\over \sqrt{g}} \de_\mu [\sqrt{g}
e^{\phi/2} f^{\mu\nu}] = 0 \ee

There are two cases to consider $\nu = \hat{\nu}$ and $\nu = u$.

For $\nu = u$ one simply gets \be \de^{\hat\mu} f_{\hat\mu u} = 0
\quad \rightarrow \quad  \de^{\hat\mu} \de_{\hat\mu} A_u - \de_u
\de^{\hat\mu} A_{\hat\mu} = 0 \ee that allows to express $A_u$ in
terms of the longitudinal component of $A_{\hat\mu}$.

For $\nu = \hat\nu$ one gets \be  \de_{\hat\mu}f^{\hat\mu\hat\nu}
+ {1\over \sqrt{g (u)}} \de_u [\sqrt{g (u)} e^{\phi/2} g^{uu}
g^{\hat\nu\hat\lambda} (\de_u A_{\hat\lambda} - \de_{\hat\lambda}
A_u)] = 0 \ee

Setting $A_{\hat\mu} = a_{\hat\mu}(u) e^{i p\cdot\hat{x}}$ and
$A_{u} = b(u) e^{i p\cdot\hat{x}}$ one can solve for
$a_{\hat\mu}(u)$ and $b(u)$. Decomposing $a_{\hat\mu}(u)$ into
longitudinal and transverse components according to \be
a_{\hat\mu}(u) = a_{\hat\mu}^T(u) + i p_{\hat\mu} a_L(u) \ee one
finds \be b(u) = a_L'(u) \ee that can be set to zero by gauge
transformations. The surviving transverse components then satisfy
an equation \be {1\over \sqrt{g (u)}} \de_u (\sqrt{g (u)}
e^{-\phi/2} \de_u a^T_{\hat\mu}) - e^{-\phi/2}p^2 a^T_{\hat\mu} =
0\ee
 that is identical to the equation for a canonical massless
scalar $\Phi$. After setting \be \Phi = e^{-\phi-g} \cY \ee  the
equation is put in canonical form with an effective potential
given by \be V_{eff} = \phi'' + g'' \approx - {1\over u^2} \quad
{\rm in \ UV} \ee Unfortunately due to the `pathological' UV
behavior there is no spectrum of discrete states associate to the
bulk massless vectors. For $SU(2)$ the story is similar. This
behavior is analogous to the one found for the fluctuations of the
metric in MN solution \cite{Maldacena:2000yy}. Indeed, the
transverse traceless components of the metric fluctuations
$h_{ij}^{TT}(u,x) = e_{ij}(p) f_p(u) e^{ipx}$ decouple from the
rest and satisfy a free massless scalar equation of the
form\footnote{The very same equation governs the dynamics of the
transverse modes of the supersymmetry partners of the graviton \eg
gravitino, graviphoton, ... as originally shown in
\cite{Bianchi:2000sm}.} \be [\de_u^2 + 4 A' \de_u - e^{2A}p^2]
f_p(u)= 0\ee

For MN solution \cite{Maldacena:2000yy}, the relevant equation has
been studied in \cite{Berg:2005pd, Berg:2006xy} and shown to have
a continuous spectrum without a mass gap. Longitudinal and radial
components of the metric mix with the active scalars and behave
better \cite{Berg:2005pd, Berg:2006xy}. Despite the area-law
behavior of the Wilson loop in MN background
\cite{Maldacena:2000yy}, the presence of massless fluctuations
casts some shadow on the holographic interpretation as a dual to a
confining theory such as $\cN=1$ SYM.

\section{Conclusions and summary}

Let us conclude by summarizing our results and draw lines for
future investigation.

We have shown that all RG flows described by the PT ansatz
\cite{Papadopoulos:2000gj} in Type IIB supergravity enjoy exact
$SU(2)\times\widetilde{ SU(2)}$ symmetry in that not only the
metric but also background $p$-forms are invariant under
diffeomorphisms generated by the six Killing vectors. We have then
identified a very general ansatz for the combined fluctuations of
metric and $p$-forms that diagonalizes the resulting equations for
the bulk massless vectors. Although derived in the context of
holography, our ansatz is expected to have much wider
applicability in any flux compactification with isometry.
Restricting our attention to the case of backgrounds invariant
under world-sheet parity $\Omega$, we have illustrated our
procedure in the case of MN solution \cite{Maldacena:2000yy} for
wrapped D5-branes. The spectrum of massless vector harmonics in
this background -- very much as the spectrum of massless scalars
and transverse traceless fluctuations of the metric
\cite{Berg:2005pd, Berg:2006xy} -- is continuous and has no mass
gap. This drawback might be related to the impossibility of fully
decoupling KK states from the desired physical modes, which
survive in the deep IR. In particular the very presence of an
exact $SU(2)\times\widetilde{ SU(2)}$ symmetry is a remnant of the
breaking of $\cN=4$ to $\cN=1^*$ with common mass for the three
chiral multiplets. Since this symmetry is an exact symmetry of any
RG flows described by PT ansatz \cite{Papadopoulos:2000gj}, not
excluding KS solution \cite{Klebanov:2000hb}, we should conclude
that holographic SYM is still undelivered \cite{Csaki:2008dt} at
least in a top-down approach. The strictly 5-d bottom-up approach
embodied by Holographic QCD \cite{Erdmenger:2007cm,
Kiritsis:2009hu, Gursoy:2009jd, dePaula:2008fp, dePaula:2009za}
 seems more promising in this
respect.

\section*{Acknowledgments}
We would like to thank C.~Bhamidipati, D.~Elander, D.~Forcella,
L.~Lopez, J.~F.~Morales, R.~Richter and M.~Samsonyan for useful
discussions. The work of M.~B. was partially supported by the ERC
Advanced Grant n.226455 {\it ``Superfields''} and by the Italian
MIUR-PRIN contract 2007-5ATT78 {\it ``Symmetries of the Universe
and of the Fundamental Interactions''}. W.~dP. acknowledges
partial support from the Brazilian agency CAPES and University of
Roma ``Tor Vergata'' for the kindly hospitality. M.~B. would like
to thank the String Theory group at Queen Mary University of
London for the kind hospitality while this work was being
completed.

\end{document}